\documentclass[pdflatex,sn-mathphys-num]{sn-jnl}

\usepackage{graphicx}%
\usepackage{multirow}%
\usepackage{amsmath,amssymb,amsfonts}%
\usepackage{amsthm}%
\usepackage{mathrsfs}%
\usepackage[title]{appendix}%
\usepackage{xcolor}%
\usepackage{textcomp}%
\usepackage{manyfoot}%
\usepackage{booktabs}%
\usepackage{algorithm}%
\usepackage{algorithmicx}%
\usepackage{algpseudocode}%
\usepackage{listings}%
\usepackage{svg}
\usepackage{caption}
\usepackage{float}

\theoremstyle{thmstyleone}%
\theoremstyle{thmstyletwo}%

\theoremstyle{thmstylethree}%

\title{Stable Long-Horizon Neural ODE Reduced-Order Models via Learned Feedback for Biological Growth and Remodeling}
\author[1]{\fnm{Joel} \sur{Laudo}}\email{jml2414@columbia.edu}

\author*[1]{\fnm{Adrian} \sur{Tepole}}\email{ab6035@columbia.edu}

\affil[1]{\orgdiv{Mechanical Engineering}, \orgname{Columbia University}, \orgaddress{\street{500 W 120th St}, \city{New York}, \postcode{10027}, \state{NY}, \country{United States}}}

\abstract{Reduced-order models (ROMs) are essential for rapid simulation of complex biomechanical systems and for bridging the gap between high fidelity models and clinical application. However, ROMs for tissue growth and remodeling (G\&R) remain largely unexplored. Here, we present a Neural Ordinary Differential Equation (NODE) ROM framework that learns latent dynamics of coupled mechanical deformation and tissue growth, demonstrated in the context of skin growth during tissue expansion (TE). TE is a challenging problem involving nonlinear contact, history-dependent material behavior, and mechanobiologically driven growth. The displacement field is compressed via Proper Orthogonal Decomposition (POD) into a low-dimensional latent space, and a NODE learns the resulting dynamics conditioned on patient-specific parameters. To address long-horizon error accumulation, a key challenge in autoregressive latent dynamical models, we propose a closed-loop architecture in which encoded features of the evolving growth field are fed back into the dynamics at each step. We compare feedback representations of increasing expressiveness: scalar, linear POD-based, and nonlinear CNN-based. The CNN-based growth feature feedback substantially stabilizes long-horizon rollouts. The best model captures 90.3\% of validation cases within clinical tolerance based on the final skin area gain, compared to 43.7\% for the open-loop baseline. Moreover, the NODE ROM achieves over 20000$\times$ the speed of full finite element simulations. More broadly, these results suggest that selectively retaining inexpensive physics of the state evolution and feeding features from these fields back into the latent dynamical system is a promising strategy for stable and accurate ROMs of G\&R in biological tissues.}

\begin{document}

\maketitle

\keywords{keyword1, Keyword2, Keyword3, Keyword4}



\section{Introduction}\label{sec1}

Tissue expansion (TE) is an important technique in reconstructive surgery in which soft tissue is gradually stretched to generate additional skin for subsequent reconstruction \cite{Neumann1957SkinExpansion}. This process is driven by the skin's mechanobiological response to sustained stretch \cite{ZOLLNER_skin_growth} and is widely used across multiple reconstructive settings, including pediatric skin repair of burn scars and congenital nevi \cite{TE_pediatric, arneja2009giant} and breast reconstruction after mastectomy \cite{radovan1982breast, TE_breast_bertozzi}, with nearly 65\% of breast reconstruction patients in the United States undergoing TE \cite{kummerow2015nationwide}.

Despite the prevalence of TE, complications remain high \cite{logiudice2003pediatric,antonyshyn1988complications}. There is a need for patient-specific pre-operative tools to inform and guide safe TE protocol design. Finite element (FE) models provide a powerful framework for simulating the nonlinear, history-dependent behavior and remodeling of soft biological tissues and organs, and have been widely used in computational biomechanics to study complex tissue responses under physiological loading \cite{tepole2011growing, MATHUR_Leaflet_remodeling}. Our past work has leveraged the theory of finite growth within continuum mechanics to model skin adaptation to stretch \cite{tepole2011growing, ambrosi2019growth}. Recently, we have extended this progress to develop patient-specific finite element models of TE and calibrated them to capture the TE deformation and growth response under prescribed clinical inflation protocols using longitudinal 3D photos from both breast and pediatric patients \cite{Acta_TE_DT, biomech_TE_DT}. These models show promise as potential patient-specific predictive tools, however the computational cost of high-fidelity TE simulation is high, particularly when repeated evaluations are required for design exploration, parameter calibration, or uncertainty quantification. This computational burden motivates the development of reduced-order models (ROMs) that can approximate FE predictions with significantly lower cost.

Traditional (projection-based) ROM methods involve the Proper Orthogonal Decomposition (POD). Discrete snapshots of governing quantities are collected from offline simulations over the input parameter range of interest and then POD is performed to generate a reduced linear basis for the solution. Typically, the governing equations are also reduced using methods such as Galerkin projection, and approximations for nonlinear terms using techniques such as the Discrete Empirical Interpolation Method can offer further computational efficiency. Such methods have proven useful for building reduced-order soft tissue models in biomechanics \cite{Soft_tissue_ROM, Aorta_Galerkin_ROM, Shah2024ROMStent}. However, these methods are intrusive and computationally complex, particularly when considered for ROMs of tissue growth and remodeling. In the context of TE, additional nonlinearities beyond nonlinear time-dependent growth and remodeling governing equations include the extremely nonlinear contact between the expander and skin. This makes the construction of efficient projection-based TE ROMs challenging and motivates the use of non-intrusive, data-driven approaches.

Recent advances in scientific machine learning have introduced neural network-based ROMs capable of learning low-dimensional representations and enabling efficient dynamical modeling of complex physical systems. For example, recent work has proposed directly learning the solution manifold using smooth neural fields and evolving the governing dynamics via Galerkin projection \cite{Jessica_Zhang_SNF_ROM}.

Alternatively, a common approach is to first learn a compressed latent representation of the physical state and then model the underlying dynamics in the latent space using Neural Operators or Neural Ordinary Differential Equations (NODEs). This compressed latent representation can be constructed using POD \cite{rojas_POD_NODE_fluidflows}, learned explicitly via models such as convolutional autoencoders \cite{Oommen2022AutoencoderDeepONet}, or learned jointly with the latent dynamics \cite{Regazzoni2024LDNet}.  Separate hybrid approaches have also been proposed to learn Koopman-inspired reduced dynamics and interpolate across parameter space, avoiding intrusive projection and large datasets \cite{DRIPS_Lu}. Latent representations have also been successfully employed in purely static settings to encode complex structure–property mappings, without explicitly modeling temporal dynamics \cite{Bouklas_latent_microstructure}. More broadly, recent work has shown that incorporating structure-informed representations into operator learning, such as learned Petrov–Galerkin weighting functions, can significantly improve generalization and data efficiency \cite{Oberai_PG_VarMiON}. Collectively, these works highlight that the choice of representation plays a critical role in the learnability and performance of data-driven models of physical systems.

However, in many cases, NODE-based ROMs can suffer from instability during long-horizon rollouts due to error accumulation, motivating mechanisms to enforce physically consistent behavior. For example, recent work has proposed stabilizing Neural ODE ROMs by embedding learned dissipative linear operators directly into the learned dynamics \cite{Stabilized_NODEs}. In addition, neural operator-based models have been shown to exhibit spectral bias and difficulty capturing sharp gradients and discontinuities, particularly in challenging settings such as fracture mechanics, further underscoring the difficulty of learning complex physical dynamics from data \cite{Lejeune_fracture_benchmark}.

Motivated by these challenges in learning stable and expressive latent dynamics, we introduce a Neural ODE ROM framework for rapid skin growth TE simulation in which features computed from the reconstructed growth state are fed back into the compressed latent dynamics as a closed-loop feedback signal. We investigate multiple feedback representations of increasing expressiveness, including a scalar summary of net area growth, low-dimensional linear features obtained via POD, and spatially resolved nonlinear features learned jointly with the dynamics using a convolutional neural network (CNN). Convolutional neural networks have been shown to effectively learn representations of spatially distributed physical fields, suggesting their suitability for extracting informative features from growth fields in this setting \cite{Garikipati_ROM_microstructure}.

The contributions of this work are summarized as follows:
\begin{itemize}
\item To our knowledge, this work is the first data-driven ROM  for full field tissue deformation and growth dynamics.
\item We showcase the approach in the context of skin growth in TE, a challenging problem due to the coupling of nonlinear dissipative material response and contact.
\item We demonstrate how long-horizon error accumulation in open-loop Neural ODE ROMs can be alleviated with a closed-loop formulation that incorporates growth feature feedback into the latent dynamics.
\item We compare multiple feedback representations, including scalar, POD-based, and CNN-based growth features, and show the CNN-based feedback significantly improves long-horizon predictive accuracy of deformation and growth.
\end{itemize}

\section{Tissue Expansion Finite Element Model}

\begin{figure}[htbp]
    \centering
    \includegraphics[width=1.0\textwidth]{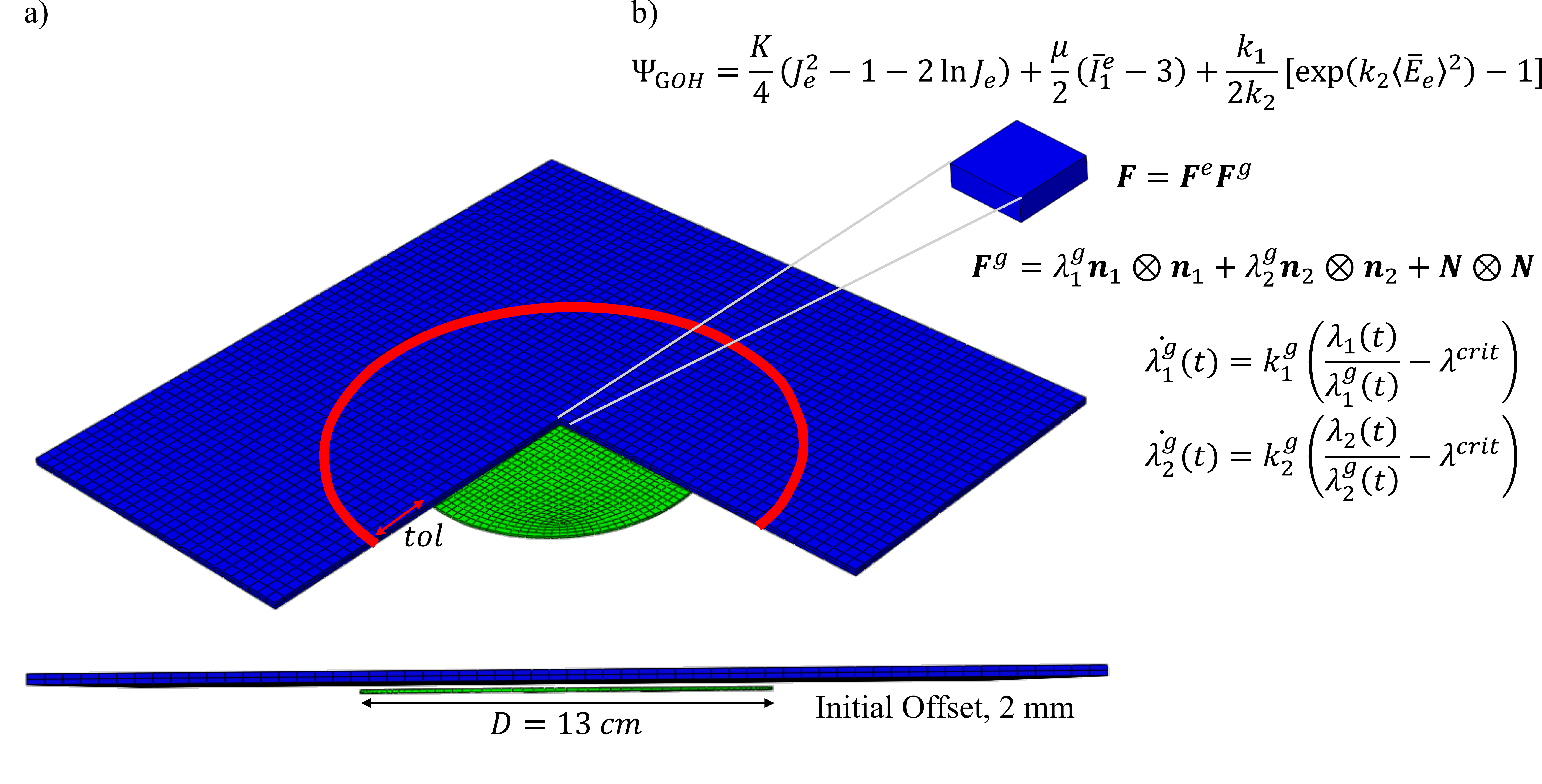}
    \caption{a) The geometry and mesh used to model skin growth in TE within the FE package Abaqus. Expander diameter $D$ and offset between expander and skin are fixed but the boundary condition radius $tol$ is sampled across simulations. b) The deformation gradient is decomposed into elastic and growth contributions and then updated at each element during each solver iteration. Parameters $\lambda^{crit},\mu,k_1^g,k_2^g,\kappa,k1$ are sampled across simulations.}
    \label{fig:FE_geom}
\end{figure}

To generate a FE dataset to construct our ROM, we simulate TE in Abaqus on a 300~mm by 300~mm idealized square skin geometry with a uniform thickness of 3~mm. The skin mesh is composed of 7200 identical C3D8 brick volume elements distributed across three  layers. We model the pre-inflated expander as a disc-shaped fluid cavity with a diameter of 13~cm, an initial cavity thickness of 1~mm and an initial separation of 2~mm from the bottom surface of the skin. We impose boundary conditions by fixing the bottom surface of the expander and also fixing all skin nodes outside of a certain radius $tol$ around the expander. We use the anisotropic hyperelastic Gasser–Ogden–Holzapfel (GOH) material model for skin material behavior \cite{Gasser2006GOH} with a strain energy function:

\begin{equation}
\Psi_{\mathrm{GOH}} =
\frac{K}{4}\left(J_e^2 - 1 - 2\ln J_e\right)
+ \frac{\mu}{2}\left(\bar{I}_1^e - 3\right)
+ \frac{k_1}{2k_2}\left[\exp\left(k_2 \,\bar{E}_e^2\right) - 1\right]
\end{equation}

We set $k_{2}=2.88$ and take the Poisson's ratio $\nu=0.45$ for near-incompressibility but sample $k_1$, $\mu$, and $\kappa$ to account for patient variability (Table \ref{tab:lhs_bounds}). We model skin growth using the framework from our past work \cite{ABT_aniso_skin_growth}. Briefly, we impose a multiplicative split of the deformation gradient, separating elastic and growth contributions:

\begin{equation}
\mathbf{F} = \mathbf{F}^{e}\mathbf{F}^{g}
\end{equation}

and then restrict growth accumulation to in-plane area change. We define the in-plane growth deformation tensor to be orthotropic:

\begin{equation}
\boldsymbol{F}^g = \lambda_1^g\, \mathbf{n}_1 \otimes \mathbf{n}_1 
     + \lambda_2^g\, \mathbf{n}_2 \otimes \mathbf{n}_2 
     + \mathbf{N} \otimes \mathbf{N}
\label{eq:Fg}
\end{equation}

with $\mathbf{n}_1$ and $\mathbf{n}_2$ the two principal fiber directions and $\mathbf{N}$ the out-of-plane normal in the reference configuration. In this work, we fix the principal fiber directions to be aligned with the $e_1$ and $e_2$ coordinate axes. We model growth accumulation by updating the principal growth stretches $\lambda_{1}^{g}$ and $\lambda_{2}^{g}$ as internal variables using the update equations:

\begin{equation}
\dot{\lambda}_{1}^{g}(t)
=
k^g_{1}\left(\lambda_1^e(t)-\lambda^{crit}\right)\, , 
\label{eq:growth_ODE_x}
\end{equation}

\begin{equation}
\dot{\lambda}_{2}^{g}(t)
=
k^g_{2}\left(\lambda_2^e(t)-\lambda^{crit}\right)\, .
\label{eq:growth_ODE_y}
\end{equation}

\begin{figure}[H]
\centering
\includegraphics[width=0.7\linewidth]{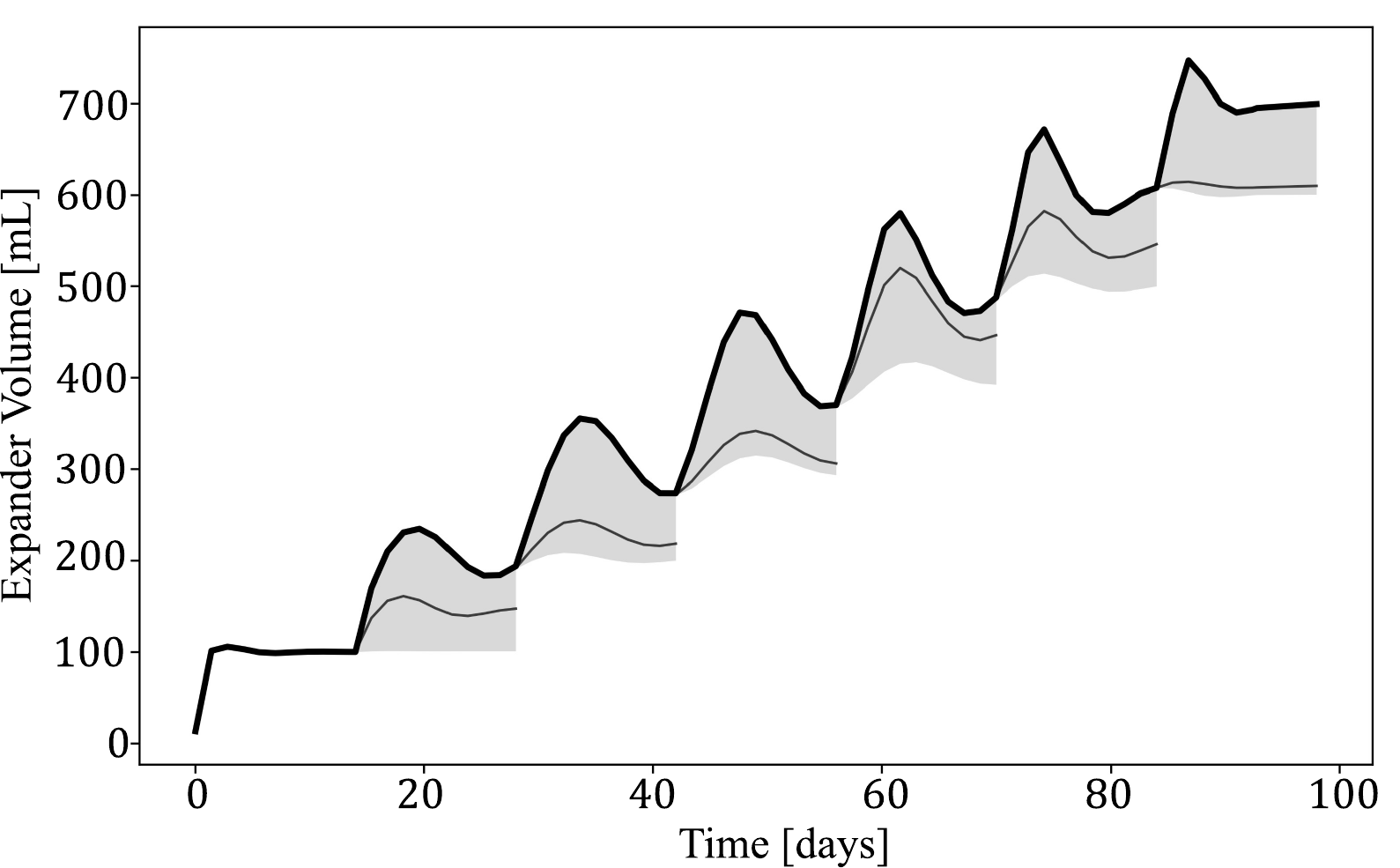}
\caption{Volume profiles from FE data. All follow a common trajectory (black) before diverging in the final stage according to the desired final volume value $V_f$. The full band of volume history variability is shown in light gray, with random samples overlaid.}
\label{fig:Vol_profiles}
\end{figure}

These equations rely on a core mechanobiological assumption which was validated in our past work \cite{Han_bayesian_calibration}, namely that the rate of change of the growth stretches is proportional to the elastic stretches in the principal fiber directions

\begin{equation}
\lambda_1^e(t)=\frac{\lambda_1(t)}{\lambda_{1}^{g}(t)}\, ,
\end{equation}

\begin{equation}
\lambda_2^e(t)=\frac{\lambda_2(t)}{\lambda_{2}^{g}(t)} \, , 
\end{equation}

minus the critical stretch $\lambda^{crit}$, via proportionality constants $k^g_1$ and $k^g_2$.  This growth update occurs element-wise over all skin elements in the geometry (Figure \ref{fig:FE_geom}b). During simulation, we impose a staged volume inflation profile to inflate the expander over time in increments of 100~mL as needed to achieve a final volume sampled between $V_f\in[100~\mathrm{mL}, 700~\mathrm{mL}]$. The volume protocol is controlled by imposing a pressure load via a proportional-integral (PI) controller with controller gains fixed to be $K_p=1.0$, $K_I=1.0$, which allows sampled volume profiles to share a common trajectory before fanning out to a specified $V_f$ (Figure \ref{fig:Vol_profiles}).

\begin{table}[htpb]
\centering
\caption{Parameter ranges used for Latin hypercube sampling.}
\label{tab:lhs_bounds}
\renewcommand{\arraystretch}{1.2}
\begin{tabular}{ll}
\hline
\textbf{Parameter} & \textbf{LHS Bounds} \\
\hline
$tol$ & $[15,\,30]\ \mathrm{mm}$ \\
$\lambda^{\mathrm{crit}}$ & $[1.0,\,1.2]$ \\
$V_f$ & $[100,\,700]\ \mathrm{mL}$ \\
$\mu$ & $[0.05,\,0.25]\ \mathrm{MPa}$ \\
$k^{g}_{1}$ & $[0.96,\,2.4]\ \mathrm{day}^{-1}$ \\
$k^{g}_{2}$ & $[0.24,\,1.2]\ \mathrm{day}^{-1}$ \\
$\kappa$ & $[0,\,0.333]$ \\
$k_{1}$ & $[1,\,10]\ \mathrm{MPa}$ \\
\hline
\end{tabular}
\end{table}

\subsection{Data Collection}
\begin{figure}[htbp]
    \centering
    \includegraphics[width=1.0\textwidth]{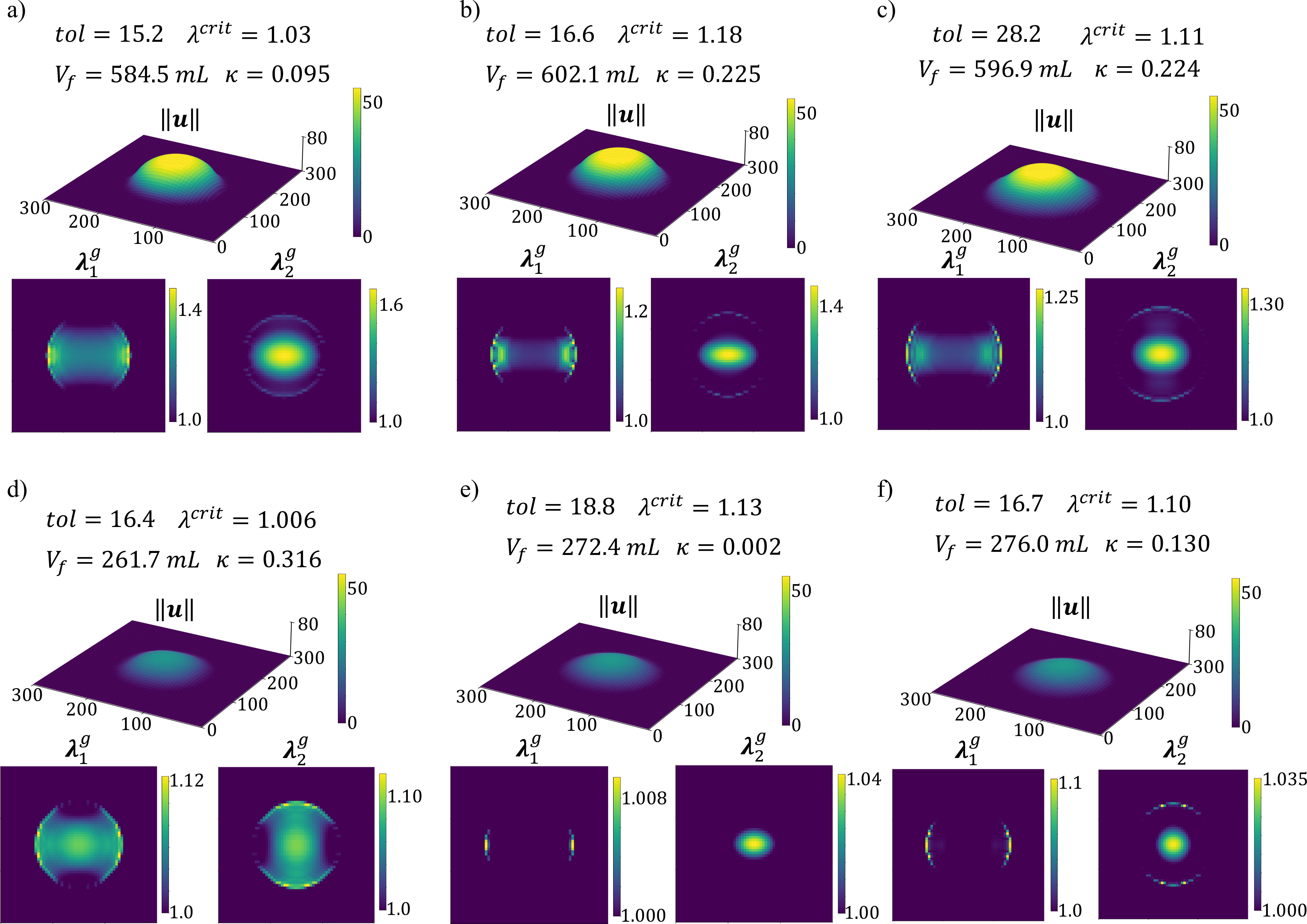}
    \caption{Representative simulation data at the final simulation timepoint.}
    \label{fig:FE_examples}
\end{figure}

We ran 1000 Latin-Hypercube sampled Abaqus simulations using the parameter ranges in Table \ref{tab:lhs_bounds} with simulation lengths that varied from 42 simulated days to 98 simulated days due to the variability in the final volume of TE considered (Fig. \ref{fig:Vol_profiles}). Of the 1000 simulations, 927 converged. Representative simulation examples are shown in Figure \ref{fig:FE_examples}. We extracted snapshots of the fluid cavity volume, displacement field, and principal growth stretch fields $\boldsymbol{\lambda}^g_1$ and $\boldsymbol{\lambda}^g_2$ from all converged simulations at a fixed time sampling increment of 1.4 days.

\begin{figure}[h!]
    \centering
    \includegraphics[width=1.0\textwidth]{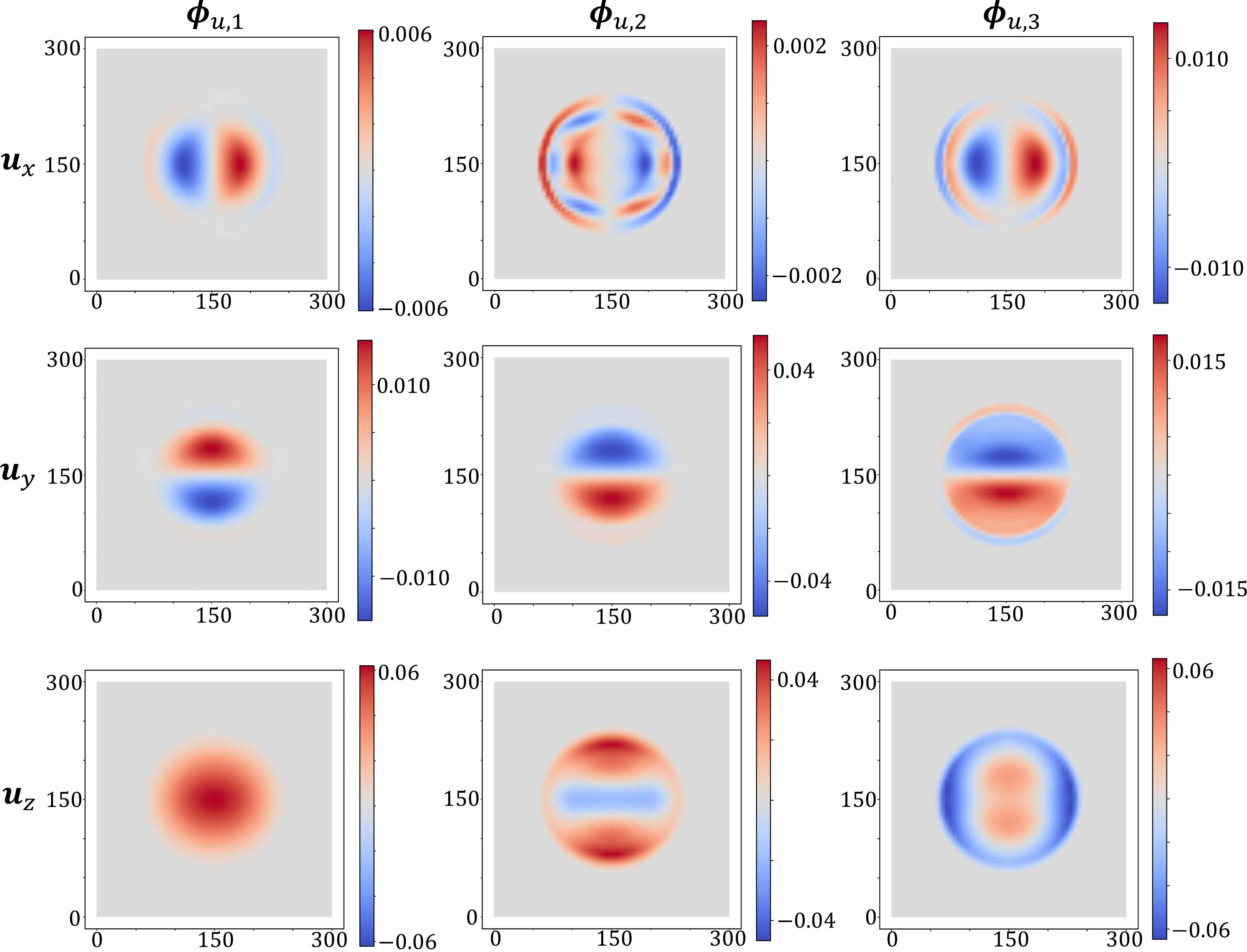}
    \caption{The first three POD modes of the displacement field snapshots across the training set split into their $\mathbf{u}_x$, $\mathbf{u}_y$, and $\mathbf{u}_z$ contributions.}
    \label{fig:POD_modes}
\end{figure}

Next we performed a train-test split on the dataset and separated the 927 converged simulations into 742 training simulations and 185 validation simulations. Using the displacement field snapshots from the training set we performed POD to generate linear basis modes for the ROM. In order to perform the POD, the displacement field was vectorized by stacking the nodal displacement components into a single column vector,
\[
\mathbf{u} = \begin{bmatrix} \mathbf{u}_x \\ \mathbf{u}_y \\ \mathbf{u}_z \end{bmatrix} \in \mathbb{R}^{3N},
\]
where $\mathbf{u}_x, \mathbf{u}_y, \mathbf{u}_z \in \mathbb{R}^N$ denote the nodal displacement components in the $x$, $y$, and $z$ directions, respectively. POD was performed on this concatenated representation, yielding a basis in $\mathbb{R}^{3N}$ whose first, second, and third portions correspond to the $x$-, $y$-, and $z$-direction displacement modes (Fig. \ref{fig:POD_modes}).

\begin{figure}[h!]
    \centering
    \includegraphics[width=1.0\textwidth]{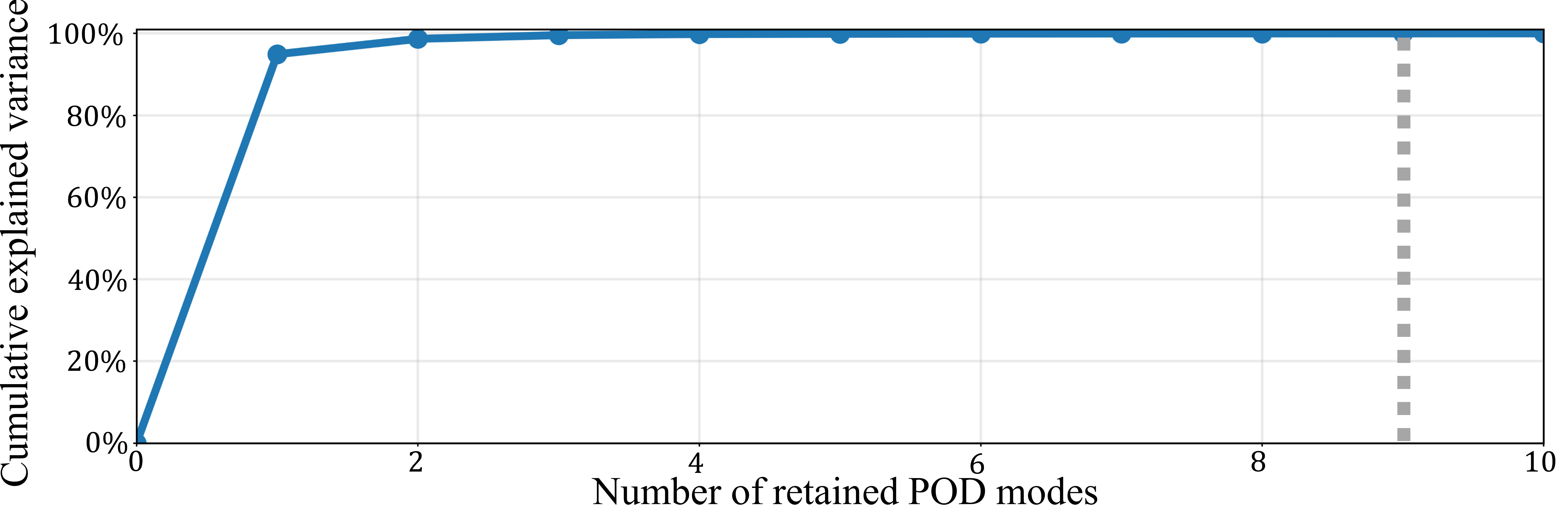}
    \caption{The cumulative explained variance plot for the POD displacement modes. We use 9 modes (dashed line) to build our NODE ROMs.}
    \label{fig:POD_cum_var}
\end{figure}

We retained 9 POD modes which corresponded to the $99.99\%$ threshold of the cumulative explained variance we specified for reconstruction of the training data (Fig. \ref{fig:POD_cum_var}). These 9 modes were used as the displacement basis for our ROM. We collected all the displacement snapshots and projected them onto the displacement basis to obtain sets of latent space trajectories. 

Let $N$ denote the number of nodes on the surface of the mesh. Let $\boldsymbol{\Phi}_{\mathbf{u}} \in \mathbb{R}^{3N \times r}$ denote the POD basis for the displacement field with $r=9$ modes, and let $\boldsymbol{\mu}_{\mathbf{u}} \in \mathbb{R}^{3N}$ denote the mean displacement field. The displacement field can then be approximated as

\begin{equation}
\mathbf{u}(t) \approx \boldsymbol{\mu}_{\mathbf{u}} + \boldsymbol{\Phi}_{\mathbf{u}} \mathbf{z}(t),
\label{eq:POD_recon}
\end{equation}
where $\mathbf{z}(t) \in \mathbb{R}^r$ are the latent coefficients. These coefficients are obtained by projection,

\begin{equation}
\mathbf{z}(t) = \boldsymbol{\Phi}_{\mathbf{u}}^T \left(\mathbf{u}(t) - \boldsymbol{\mu}_{\mathbf{u}}\right).
\label{POD_project}
\end{equation}

Therefore, the latent dimension of $r=9$ is much smaller than the full field dimension for the displacement field of $3N=11163$ degrees of freedom.

\section{Constructing Reduced-Order Dynamical Surrogates}

 Using the POD basis constructed from our training data displacement snapshots, we can reconstruct the displacement at any timepoint using the POD projection matrix and the vector of 9 POD coefficients using Equation \ref{eq:POD_recon}. These 9 POD coefficients serve as the reduced-order latent state for the displacement. Since the displacement is kinematically driven by the scalar expander volume, we treat the 9 POD displacement coefficients and the fluid cavity volume together as a complete reduced-order state for the system and thus attempt to learn the dynamics for the augmented latent state

 \begin{equation}
\tilde{\boldsymbol{z}}=[\boldsymbol{z},\, V] \in \mathbb{R}^{10}
\label{eq:aug_state}
 \end{equation}
 using a Neural Ordinary Differential Equation (NODE). The NODEs are simple fully connected networks with two hidden layers and are conditioned on simulation input parameters that include boundary conditions $tol$, material properties $\kappa$, $k_1$, $\mu$, growth model parameters $\lambda^{crit}$, $k^g_1$ and $k^g_2$ and volume PI controller terms which are updated each iteration. We use this model as a learned representation of the latent dynamical system to test four architectures:
 
\begin{enumerate}
\item Open-loop NODE without growth feedback (\textbf{Model A})
\item NODE with scalar net area gain feedback (\textbf{Model B})
\item NODE with PCA growth features (\textbf{Model C})
\item NODE with CNN-based growth features (\textbf{Model D})
\end{enumerate}

\subsection{Open-Loop NODE Predictor}

\begin{figure}[htbp]
    \centering
    \includegraphics[width=1.0\textwidth]{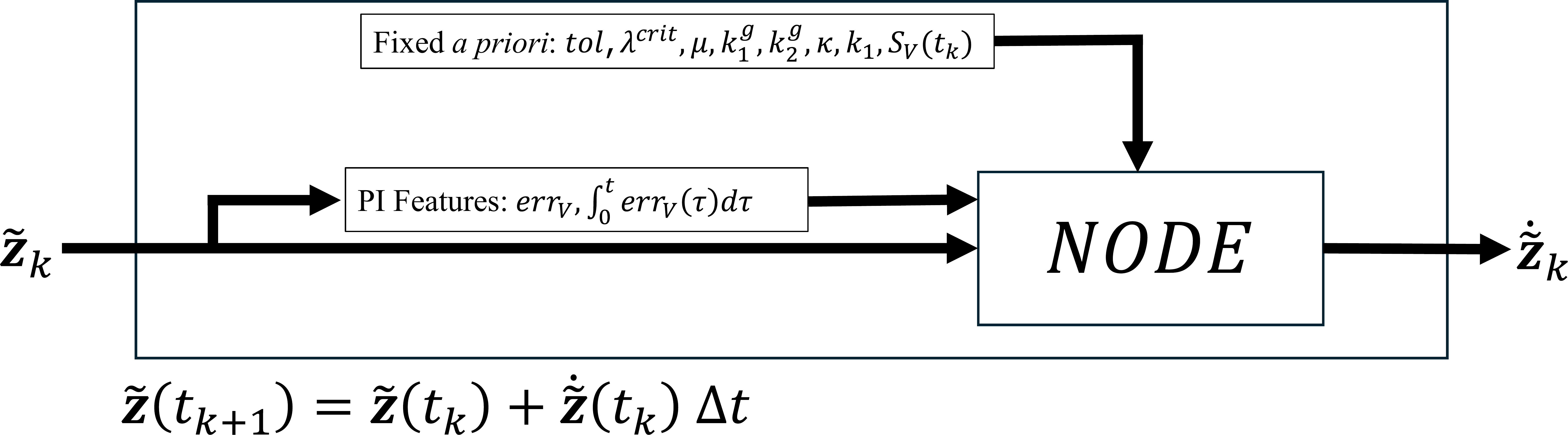}
    \caption{The open-loop predictor for system latent dynamics (\textbf{Model A}). We supervise and rollout the NODE  using a fixed timestep to match our FE data.}
    \label{fig:OL_Predictor_Arch}
\end{figure}

\begin{figure}[htbp]
    \centering
    \includegraphics[width=1.0\textwidth]{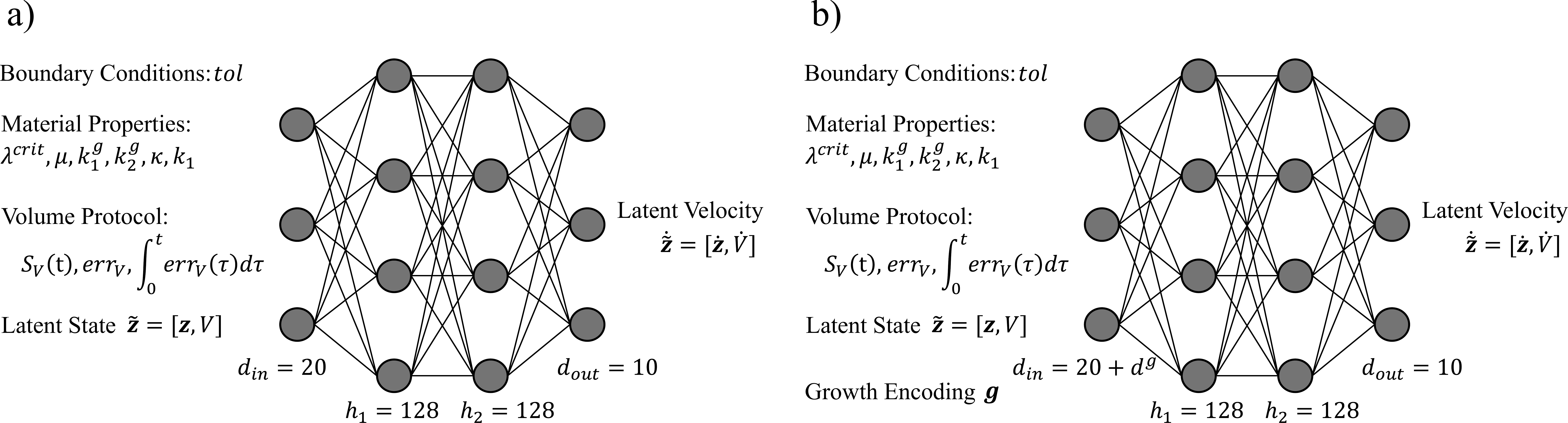}
    \caption{The neural network architectures for a) the open loop predictor and b) the closed-loop predictors}
    \label{fig:MLP_arch}
\end{figure}

The open-loop NODE-ROM, which we will call Model A, is shown in Figure \ref{fig:OL_Predictor_Arch}, and it uses the network architecture in Figure \ref{fig:MLP_arch}a.

Here, the input of the NODE includes the boundary condition variable $tol$, the three properties of the material ($\kappa$, $k_1$ and $\mu$ sampled from the GOH model), and the three skin growth model parameters ($\lambda^{crit}$, $k^g_1$ and $k^g_2$). These variables take fixed values for each simulation and are held constant during open-loop rollout for a single simulation. 

We also pass additional terms derived from the current volume, volume setpoint, and volume history. In our Abaqus UMAT, the expander volume is controlled indirectly by prescribing a fluid pressure using a PID controller. This means that to ensure that the learned volume dynamics are Markovian, the network requires complete information about the controller effort at each prediction step. This information includes the current volume setpoint $S_{V}(t_k)$, which is known given the current time $t_k$, the current error in the volume $err_{V}=S_{V}^{(k)}-V$, and the integrated time history of the error $\int_0^t \mathrm{err}_{V}(\tau)\, d\tau $. These terms are recomputed at each prediction step from the current volume state and the saved volume state history. 

Given these inputs and the latent state $\tilde{\boldsymbol{z}}$, the NODE predicts the latent state velocity $\dot{\tilde{\boldsymbol{z}}}$. This velocity is then integrated using Forward Euler with a constant time step $\Delta t=1.4~days$ using the update:

\begin{equation}
\tilde{\boldsymbol{z}}(t_{k+1}) = \tilde{\boldsymbol{z}}(t_k) + \dot{\tilde{\boldsymbol{z}}}(t_k)\,\Delta t
\end{equation}

This Forward Euler scheme was chosen to match the saved FE data. We can integrate the model forward in time to produce latent state trajectores and then decode the latent displacement vector at each step using the POD basis to produce the full displacement field. We can run this open-loop model forward using the update equations for many steps until we have generated an entire trajectory.

\subsection{Closed-Loop NODE Predictors with Growth Feature Feedback}

Anticipating model drift from open loop tissue growth ROMs, we introduce a closed-loop NODE predictor architecture that modifies the open-loop predictor with a feedback signal to the NODE from the current growth state. This structure more closely mimics the coupled FE solver update for mechanics and growth than the open-loop predictor, which evolves the displacement without any embedded growth information. To create this structure, we modify the NODE input structure  to pass an additional input, a growth encoding $\mathbf{g}$ with dimensionality $d_g$ (Figure \ref{fig:MLP_arch}b).

\begin{figure}[htbp]
    \centering
    \includegraphics[width=1.0\textwidth]{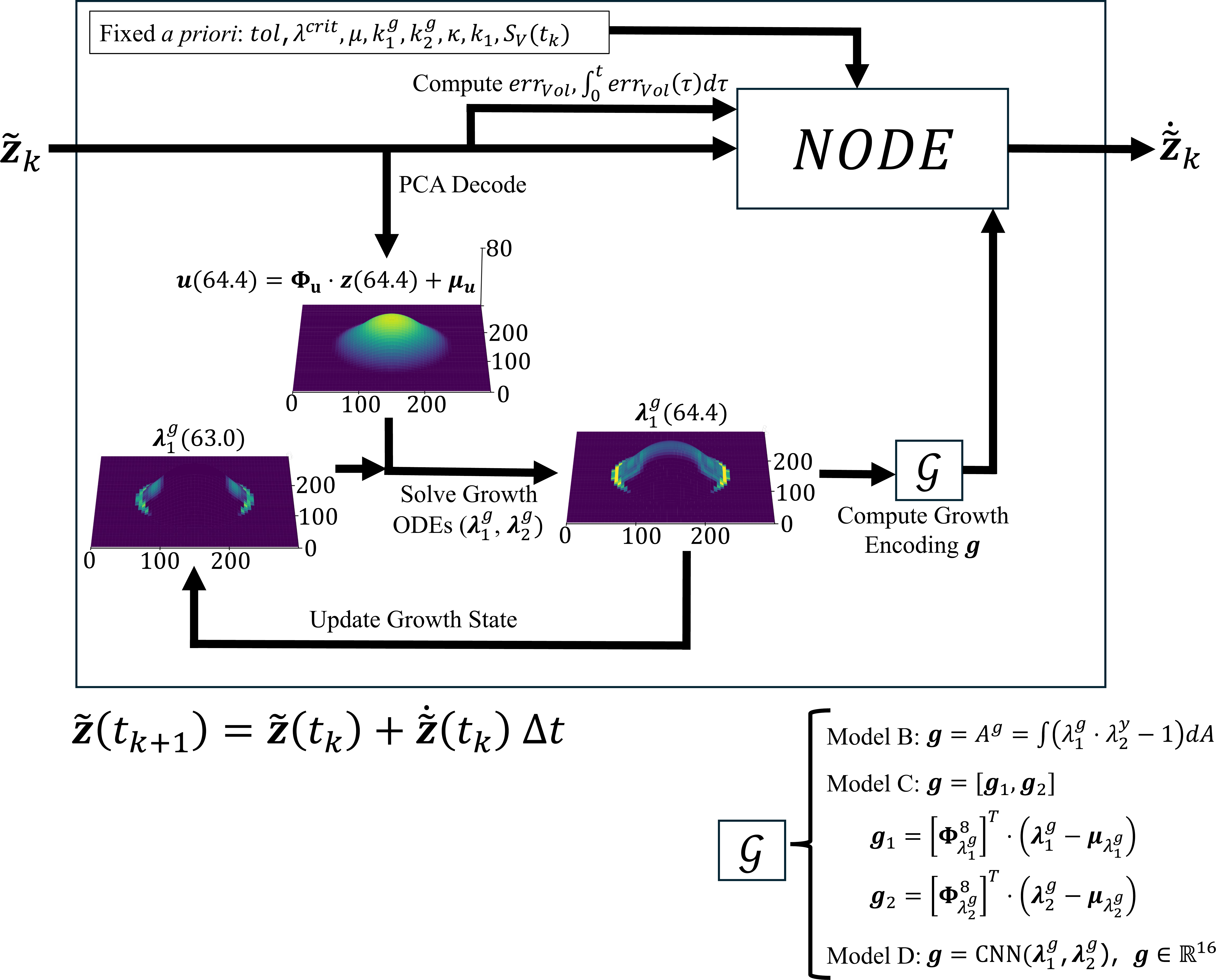}
    \caption{The closed-loop predictor architecture. The latent state is decoded into the full displacement field which is used to update the stored growth fields. A growth field operator $\mathcal{G}$ encodes information from these fields which is then passed to the Neural ODE as a closed-loop feedback signal.}
    \label{fig:CL_Predictor}
\end{figure}

The closed-loop architecture is detailed in Figure \ref{fig:CL_Predictor}. At the start of the forward iteration, the current reduced displacement state $\mathbf{z}_k$ is decoded via the POD basis into the full displacement field $\mathbf{u}_k$. Using this displacement field, the growth fields $\boldsymbol{\lambda}_1^g$, $\boldsymbol{\lambda}_2^g$ from the previous iteration are updated by solving Equations \ref{eq:growth_ODE_x} and \ref{eq:growth_ODE_y} over the geometry. The updated growth fields are passed to an operator $\mathcal{G}$ which computes an encoded representation of the growth fields $\boldsymbol{g}$ that is passed to the NODE as a closed-loop feedback signal.  

We choose three different encoding representations $\boldsymbol{g}$ and train three different models which we call Model B, Model C, and Model D. Model B encodes growth information as a single scalar, the net integrated area gain:

\begin{equation}
A^g = \int \left( \lambda_1^g \cdot \lambda_2^g - 1 \right)\, dA
\label{eq:Ag_feedback}
\end{equation}

which is normalized using training data statistics before being passed to the model. The encoding for model C uses POD bases for the $\lambda_1^g$ and $\lambda_2^g$ fields constructed from the training data. At each iteration, the updated growth fields $\lambda_1^g$ and $\lambda_2^g$ are projected onto their respective linear manifolds and the coefficients are stored as latent representations $\boldsymbol{g}_1$ and $\boldsymbol{g}_2$ and then concatenated into the final embedding $\boldsymbol{g}=[\boldsymbol{g}_1, \boldsymbol{g}_2]$ which is then normalized using training data statistics and passed to the NODE. Model D uses a lightweight CNN encoder with four $3 \times 3$ convolutional layers with Group Normalization \cite{wu2018groupnormalization} and SiLU activations. The convolutions were strided for progressive spatial downsampling and channel expansion. The final feature map was globally pooled and projected to a low-dimensional embedding $\boldsymbol{g} \in \mathbb{R}^{16}$. 

\subsection{Training Scheme}\label{sec:train-scheme}

We define the one-step loss in latent space as the mean squared error (MSE) between the current and next latent state over all pairs in the dataset:

\begin{equation}
\mathcal{L}_{\mathrm{1step}}
=
\frac{1}{N_b}
\sum_{i=1}^{N_b}
\left\|
\tilde{\mathbf{z}}_{k+1}^{(i),\mathrm{pred}}
-
\tilde{\mathbf{z}}_{k+1}^{(i),\mathrm{true}}
\right\|_2^2
\label{eq:one_step_loss}
\end{equation}

Note that the latent state used here is the augemented latent state (normalized) with both the latent displacement coefficients and the expander volume.

We define a tail loss which is our long-horizon training objective. We penalize the average smooth $L_1$ loss, which we denote $\mathcal{D}_{1s}$, over the normalized latent displacement coefficients in the last 0.30 fraction of each rollout as this tail loss:  

\begin{equation}
\mathcal{L}_{\text{tail}} = \frac{1}{\sqrt{T}} \cdot \frac{1}{T - T_0} \sum_{k = T_0}^{T-1} \mathcal{D}_{1s}\!\left(\boldsymbol{z}_k^{\text{pred}},\, \boldsymbol{z}_k^{\text{true}}\right)\, ,
\label{eq:tail-loss}
\end{equation}

where $T$ is the length of the simulation rollout in steps and $T_0=0.70\cdot (T-1)$. Here we correct for the random walk bias implicit in Forward Euler rollouts by including a factor $\frac{1}{\sqrt{T}}$ for reasons we make clear next.

\subsubsection{Scaling of Tail Loss via Random Walk Error Accumulation}

In the rollout setting, prediction errors at each time step accumulate as the model is iteratively applied. To motivate the normalization used in the tail loss, we analyze the statistics of this accumulation under mild assumptions.

Consider the Neural ODE (NODE) update in latent space:
\begin{equation}
\tilde{\mathbf{z}}_{k+1} = \tilde{\mathbf{z}}_k + f_\theta(\tilde{\mathbf{z}}_k)\Delta t.
\end{equation}

Let the prediction error at step $k$ be defined as
\begin{equation}
\mathbf{e}_k = \tilde{\mathbf{z}}_k^{\,\text{pred}} - \tilde{\mathbf{z}}_k^{\,\text{true}}.
\end{equation}

Over a rollout of length $T$, the cumulative error can be expressed as
\begin{equation}
\mathbf{E}_T = \sum_{k=1}^{T} \mathbf{e}_k.
\end{equation}

We assume that the per-step errors $\{\mathbf{e}_k\}$ have zero mean, are approximately uncorrelated, and have comparable variance, i.e. $\mathrm{Var}(\mathbf{e}_k) \approx \sigma^2$.

Under these assumptions, the cumulative error behaves like a random walk. In particular,
\begin{equation}
\mathrm{Var}(\mathbf{E}_T) = \sum_{k=1}^{T} \mathrm{Var}(\mathbf{e}_k) \approx T \sigma^2.
\end{equation}

Taking the root-mean-square (RMS) magnitude yields
\begin{equation}
\sqrt{\mathbb{E}\left[\|\mathbf{E}_T\|^2\right]} \sim \sqrt{T}\,\sigma.
\end{equation}

Thus, even if the per-step error statistics remain constant, the accumulated rollout error grows proportionally to $\sqrt{T}$.

The tail loss is constructed by averaging errors over the final portion of the rollout, with the factor $1/\sqrt{T}$ compensating for the $\mathcal{O}(\sqrt{T})$ growth in accumulated error due to random walk statistics. Without this normalization, longer rollouts would systematically incur larger losses purely due to stochastic accumulation, rather than reflecting a degradation in per-step model accuracy.

Therefore, the normalization ensures that the tail loss remains comparable across different rollout horizons and more accurately reflects the underlying quality of the learned dynamics, rather than the length of the rollout.

\subsubsection{Growth Loss Guardrail Penalty and Training Curriculum}

We also define a loss penalty on the final integrated scalar area gain $A^{g}_{f}$:

\begin{equation}
\mathcal{L}_{Ag} = \left(
\frac{A^{g}_{f,\text{pred}} - A^{g}_{f,\text{true}}}
{A^{g}_{f,\text{true}} + 10~\text{cm}^2}
\right)^2
\label{eq:Ag_loss}
\end{equation}

This $A^{g}_{f}$ loss is a penalty is on the relative error in the final integrated area gain with an offset of $10~\text{cm}^2$ in the denominator to prevent huge errors near $A^{g}_{f,\text{true}}=0$. Penalizing the relative error ensures that the penalty is meaningful for simulations with a wide range of $A^{g}_{f,\text{true}}$ values. This quadratic penalty acts as a guardrail that discourages catastrophic rollouts in the physical displacement space. We combine this penalty with the tail-loss which then serves as the total rollout loss for predicted simulations. Since long-horizon training rollouts can lead to catastrophically large loss gradients, we impose a cap on the rollout loss so that any rollout loss gradients larger than the cap are attenuated to the cap magnitude without affecting the gradient direction. This prevents catastrophic long-horizon training predictions from destabilizing training while still allowing them to influence training. We define the rollout loss as the average of the combined tail-loss $\mathcal{L}_{\text{tail}}$ and $A^{g}_{f}$  for all long-horizon training rollouts in the batch $N_{sims}$, where the average is taken after the cap $c$ is applied to the combined loss for each simulation:

\begin{equation}
\mathcal{L}_{\text{roll}}
=
\frac{1}{N_{\text{sims}}}
\sum_{i=1}^{N_{\text{sims}}}
\mathrm{max}\!\{
\mathcal{L}_{\text{tail},i}
+
\lambda_{A^g}\,\mathcal{L}_{A^g,i}
, c\} \, .
\label{eq:rollout_loss}
\end{equation}

In practice, we found that a cap of $c=0.033$ was appropriate. The rollout loss was scaled by a hyperparameter $\lambda_{roll}=1.8E-3$ when computing the total loss accounting for 1-step and tail contributions. 

All four models were trained for 360 total epochs. The first 40 epochs were warmup epochs during which we only train using Equation \ref{eq:one_step_loss}. Then during all subsequent epoch we solve a 2-stage optimization problem. We first train on all one-step training pairs in the dataset using Equation \ref{eq:one_step_loss} and update the model parameters. Then we sample a batch of training simulations and compute the long-horizon model rollout prediction loss on this batch using Equation \ref{eq:rollout_loss} and update the model parameters again. This separated training strategy means that for each epoch we first satisfy the local dynamics and then smooth these dynamics globally to reduce long-horizon errors.

After the warmup epochs, we sample long-horizon training simulations according to a learning curriculum shown in Table \ref{tab:volume_curriculum}, where simulations with progressively larger final volumes are introduced. Since longer rollouts share the early-stage volume dynamics of shorter ones, this curriculum allows the model to first learn stable short-horizon behavior before incrementally extending to longer horizons, while maintaining previously learned dynamics. Additionally, introducing long rollouts too early leads to large accumulated errors and unstable training, so this staged approach improves optimization by avoiding catastrophic loss growth in the early epochs.

\begin{table}[ht]
\centering
\caption{Volume curriculum schedule used during rollout training.}
\label{tab:volume_curriculum}
\renewcommand{\arraystretch}{1.3}
\begin{tabular}{c|c|c|c|c|c|c|c}
\hline
\textbf{Volume Stage} & \textbf{1} & \textbf{2} & \textbf{3} & \textbf{4} & \textbf{5} & \textbf{6} & \textbf{7} \\
\hline
\textbf{$V_{cap}$} & 200 cc & 300 cc & 400 cc & 500 cc & 600 cc & 700 cc & 700 cc \\
\hline
\textbf{Frac New} & N/A & 0.45 & 0.40 & 0.35 & 0.35 & 0.30 & Uniform \\
\hline
\textbf{\# Epochs} & 40 & 40 & 40 & 40 & 40 & 60 & 60 \\
\hline
\end{tabular}
\end{table}

We separate all training simulations into bins based on the final volume in increments of $100~cc$ and then sample from these bins according to a volume cap which is increased as the stage progresses. During each stage we sample a fraction ('$\mathrm{Frac\ New}$' in Table \ref{tab:volume_curriculum}) of simulations from the bin with $V_f\in[(V_{cap} -100), V_{cap}]~cc$ and the remaining fraction of simulations from all previous bins which have $V_f\in[100, (V_{cap} -100)]~cc$. During the final volume stage, we sample equally from all bins. We use random sampling without replacement so that training simulation rollouts are not repeated before we have cycled through all examples and sample 40 rollouts per epoch which is equivalent to approximately 17.25 cycles through the training simulation dataset. However, due to the volume learning curriculum this average is not evenly distributed as simulations with $V_f\in[100, 200]~cc$ are seen about 35 times each while simulations with $V_f\in[600, 700]~cc$ are seen about 9 times each. Overall, this curriculum emphasizes repeated exposure to short-horizon dynamics while gradually introducing longer rollouts, enabling stable training by limiting early error accumulation and allowing the model to extend to longer horizons without forgetting previously learned behavior.

\section{Results}\label{sec2}

As described in the Methods section, a Latin hypercube sampling design of 1000 simulations was generated using the bounds shown in Table \ref{tab:lhs_bounds}. Of these 1000 Abaqus simulations, 927 converged and were separated into 742 training simulations and 185 validation simulations. 
We trained all four models according to the training scheme detailed in Section \ref{sec:train-scheme} for 360 total epochs, using the same random seed and the same hyperparameter values of $\lambda_{roll}=1.8\cdot10^{-3}$, $\lambda_{A^g}=0.082$ and $c=0.033$.

\begin{figure}[htbp]
    \centering
    \includegraphics[width=1.0\textwidth]{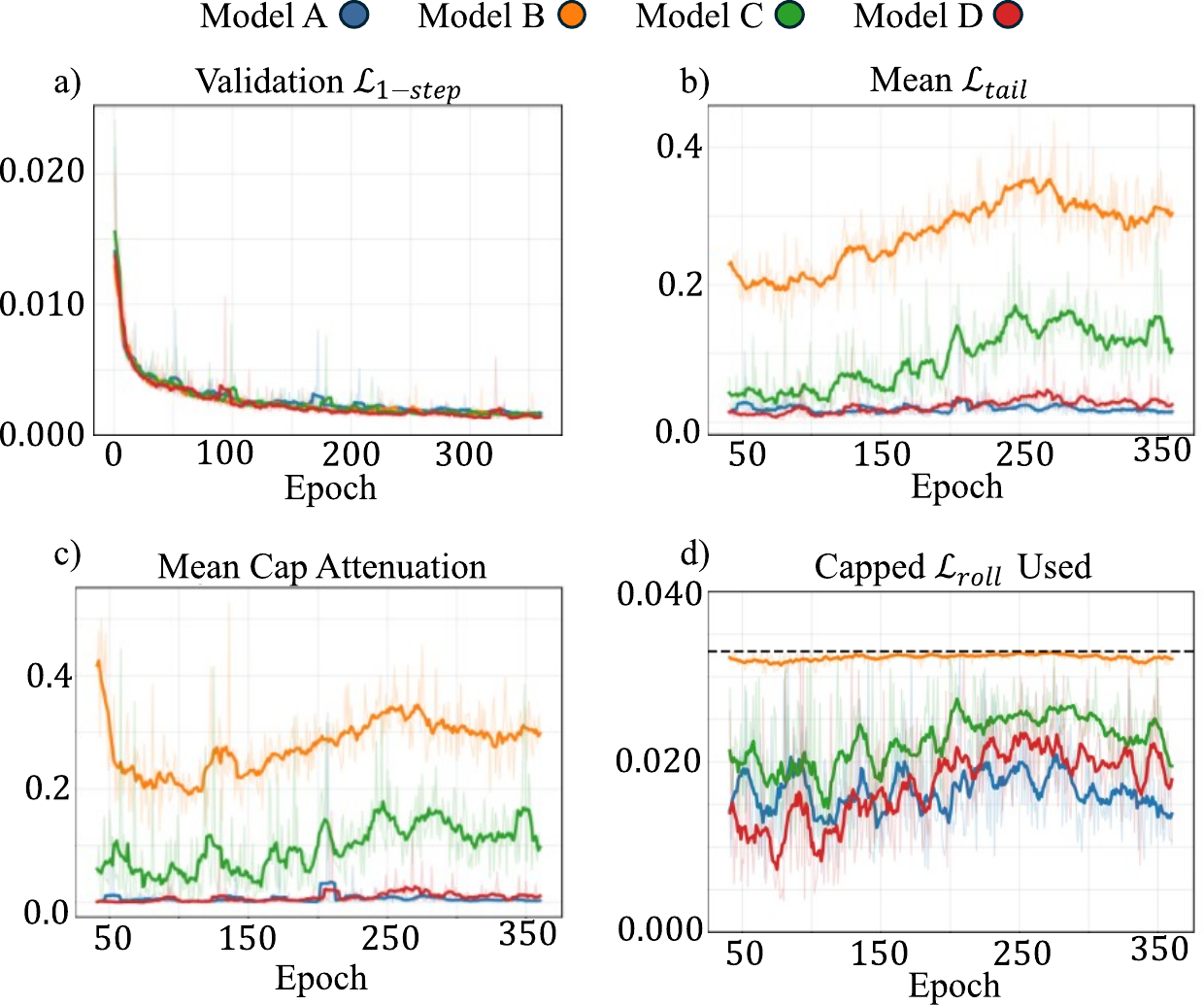}
    \caption{Training diagnostic plots across the four models with window averaging for easier visualization. The raw curves are shown faintly in the background. a) All models achieve similar validation one-step losses. b) Model A and Model D both have low tail losses while Model B and C have high tail losses during training. These high losses require high cap attenuation c), defined as the difference between the mean pre-cap rollout loss and the mean post-cap rollout loss, to ensure training remains stable. d) The capped rollout loss used during training is shown, Model B hits the cap consistently.}
    \label{fig:Train_diagnostics}
\end{figure}

Training diagnostic plots are shown in Figure \ref{fig:Train_diagnostics}. All four models achieve similar one-step losses on the training and validation datasets. We saved the models with the best validation one-step losses which are reported in Table \ref{tab:best_val_loss}. The models differ in the tail loss, with models A and D, which were the open-loop and CNN feedback configurations respectively, achieving the lowest tail losses. Model C, based on growth POD feedback, showed the third highest tail loss, with model B, using a single area scalar as feedback, showing the worst performance. 

\begin{table}[ht!]
\centering
\caption{Best validation one-step loss for each saved model.}
\label{tab:best_val_loss}
\renewcommand{\arraystretch}{1.2}
\begin{tabular}{lcc}
\hline
\textbf{Model} & \textbf{Loss} & \textbf{Epoch} \\
\hline
A & $1.56 \times 10^{-3}$ & 345 \\
B & $1.29 \times 10^{-3}$ & 325 \\
C & $1.39 \times 10^{-3}$ & 358 \\
D & $1.14 \times 10^{-3}$ & 333 \\
\hline
\end{tabular}
\end{table}


With the trained models, we can evaluate rollout predictions of full field displacements and growth fields. An example for the open-loop model A is shown in Figure \ref{fig:OL_disp_growth}. The trajectories are evaluated with the NODE in the reduced space $\tilde{\mathbf{z}}\in \mathbb{R}^{10}$. However, because of the linear POD basis, the full field displacement, $\mathbf{u}\in\mathbb{R}^{11163}$, can be easily reconstructed. The growth field update is always done in the physical space, that is $\boldsymbol{\lambda}_1^g, \boldsymbol{\lambda}_2^g \in \mathbb{R}^{3600}$. In the case of open-loop dynamics of model A, the growth update is a postprocessing step and it does not feed into the latent dynamics, unlike models B, C, and D which have the growth feedback illustrated in Figure \ref{fig:CL_Predictor}.

\begin{figure}[htbp]
    \centering
    \includegraphics[width=1.0\textwidth]{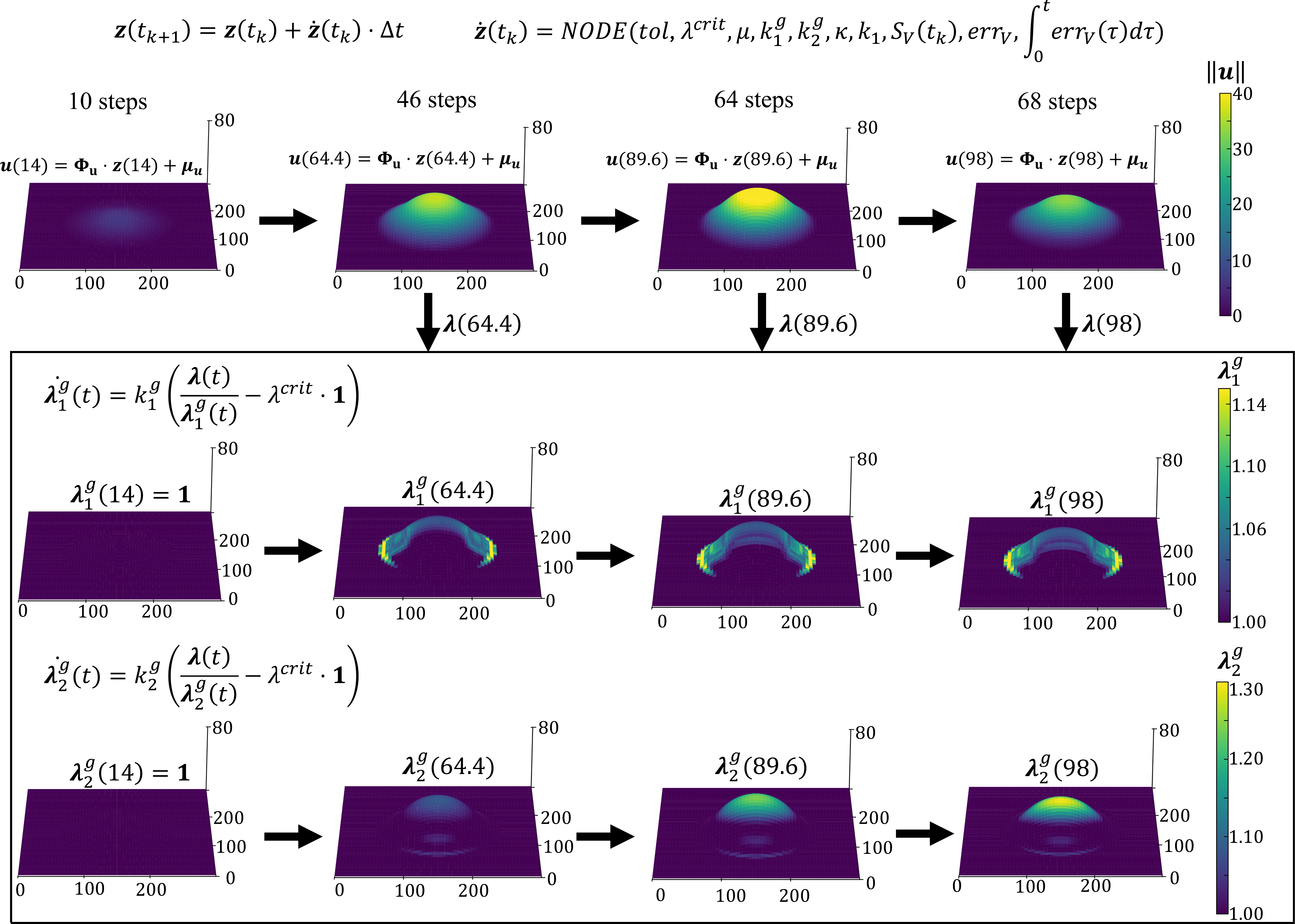}
    \caption{Example open-loop autoregressive rollout. At each step we can evolve the surface growth ODEs outside the loop to progress the growth state.}
    \label{fig:OL_disp_growth_steps}
\end{figure}

\begin{figure}[htbp]
    \centering
    \includegraphics[width=1.0\textwidth]{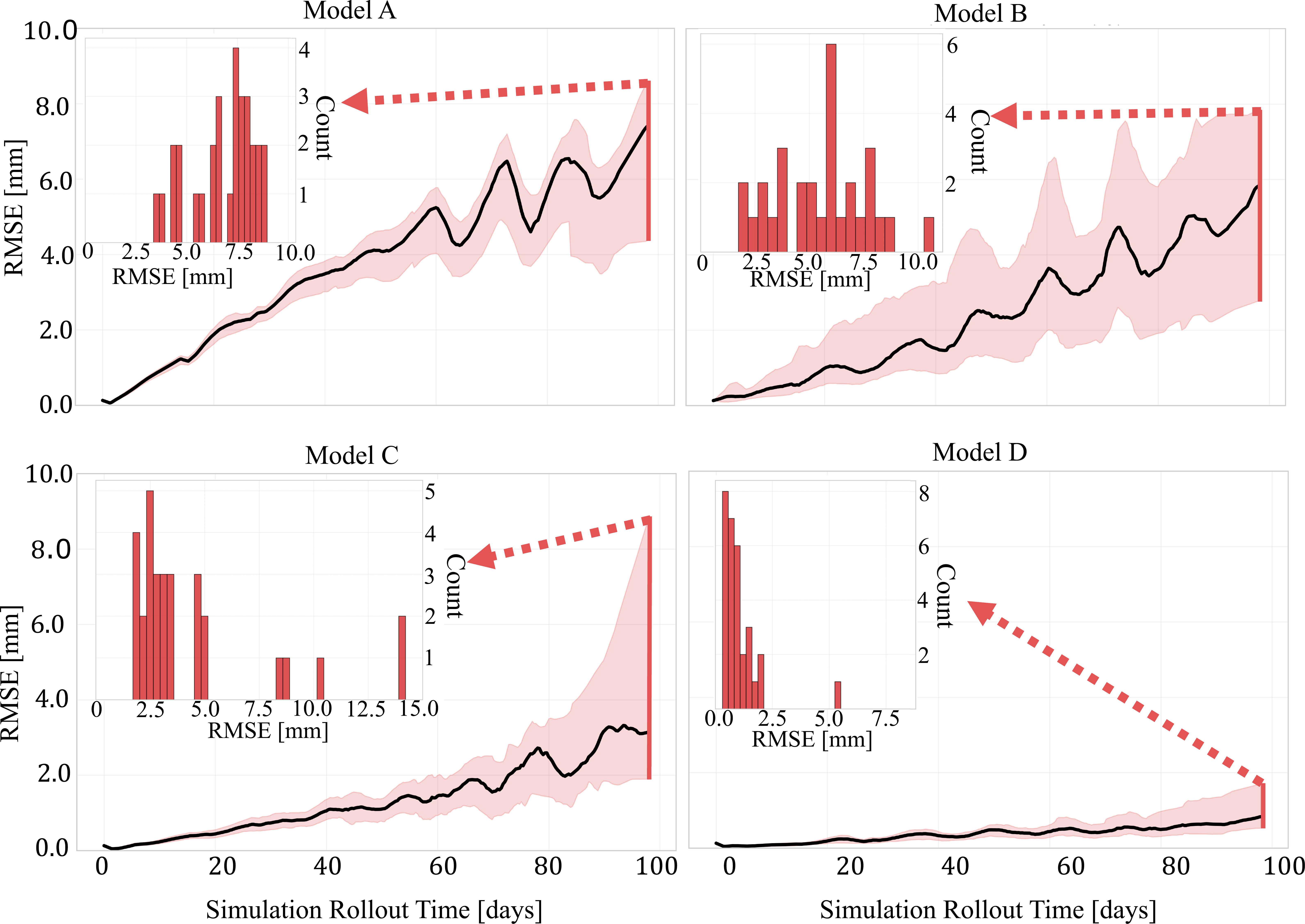}
    \caption{Reconstructed displacement RMSE [mm] vs. simulation rollout time across the four models on 185 validation simulations. The black curve shows the median RMSE while the red envelope captures the 10\% and 90\% performance interval. Note that not all validation simulations have the same lengths. The red histograms show the distribution of errors for the predictions that reach the final rollout time of 98.0 days.}
    \label{fig:Disp_recon}
\end{figure}

We evaluated the performance of all four models on 185 validation simulations by rolling the models out autoregressively and then comparing the reconstructed displacement at each timestep against the true displacement (Figure \ref{fig:Disp_recon}). 

While the one-step latent losses are similar across models, the displacement reconstruction accuracy varies greatly between them (Figure \ref{fig:Disp_recon}). In particular, Model D outshines the rest, demonstrating that flexible growth feature feedback can enable stable and accurate closed loop displacement dynamics. Model B and Model C perform better on the displacement reconstruction in the median case than Model A, but Model B has a much higher variance with some displacement trajectories significantly improved and some worsened depending on the usefulness of the feedback signal $A^g$. Model C, based on growth POD features, in particular shows signs of prediction instability as time progresses. Note that simulations have differing rollout endtimes with simulations with longer rollout endtimes corresponding to higher final volumes, and so as time progresses, the model has fewer simulation examples to learn from. This is likely a contributing factor in the sharp increase in Model C's reconstruction error in the last 20 rollout days, which is the region in which the model has the fewest training examples. This contribution is also likely why the error bounds for Model D grow larger in the last 20 rollout days. A future clinically usable surrogate would either be given more training examples in this region, or restricted to a usable range on the interior of its final volume domain, say restricting $V_f\leq600~cc$.

\begin{figure}[htbp]
    \centering
    \includegraphics[width=1.0\textwidth]{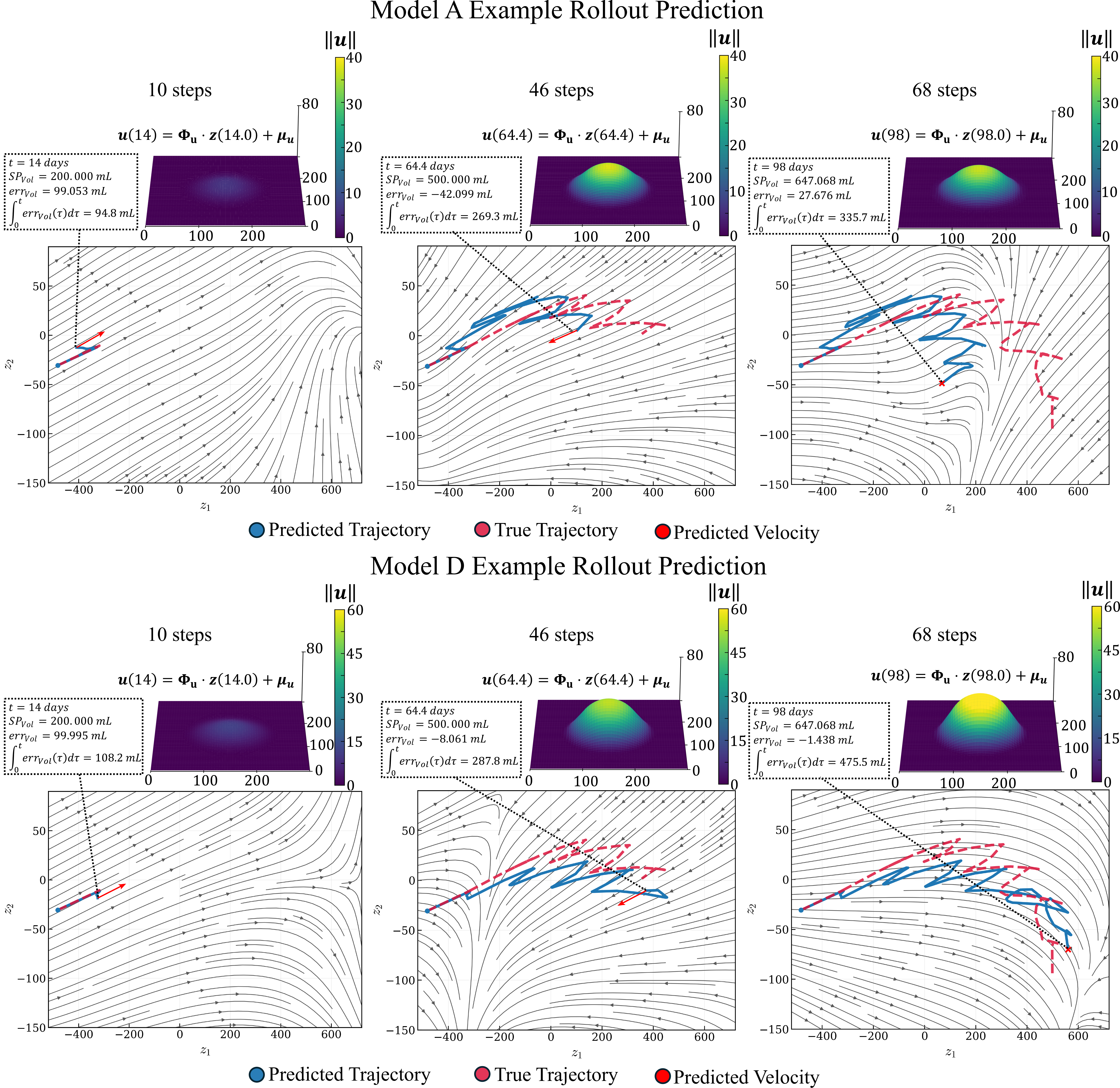}
    \caption{Model D tracks the dominant latent mode with much higher accuracy than Model A (representative validation case). Black streamlines show the NODE vector field conditioned on simulation inputs and current values of $S_V$, $err_V$, and $\int_0^t \mathrm{err}_V(\tau),d\tau$, with additional dependence on CNN features for Model D (not shown).}
    \label{fig:OL_Rollout}
\end{figure}

A clearer contrast between the open-loop Model A and the CNN-growth feedback in Model D is depicted in Figure \ref{fig:OL_Rollout}. The trajectories in a slice of the latent space are shown for the two models, together with the full field reconstruction from POD. The true trajectory in latent space for the same validation case is also shown, to allow direct comparison against the predicted trajectory from models A and D.  The true trajectory in latent space shows the extreme nonlinearity that is expected given that the process is governed by a PID volume controller, contact between skin and expander, and nonlinear mechanics with dissipation due to growth. The open-loop NODE drifts away from the true trajectory as the rollout time increases. In contrast, the feedback using CNN-features from the growth field result in accurate latent trajectories for very long rollout times.

\begin{figure}[htbp]
    \centering
    \includegraphics[width=1.0\textwidth]{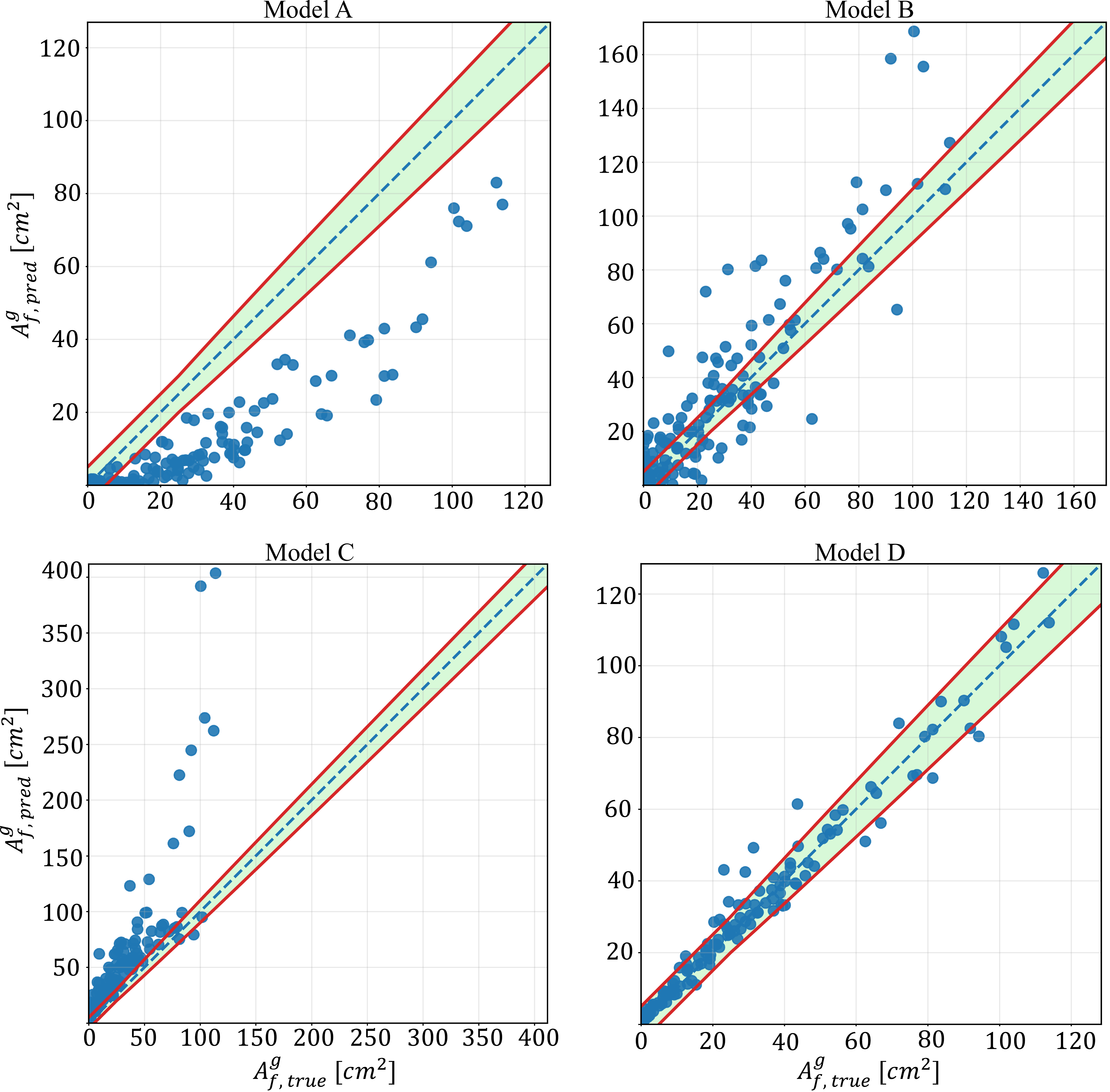}
    \caption{Validation set performance on the final integrated scalar area gain vs the true scalar area gain across models. Only Model D successfully predicts the true final area gain with accuracy.}
    \label{fig:Ag_recon}
\end{figure}

We also compare the performance of the four models at predicting the final scalar integrated area gain $A^g_{f,pred}$ on the validation set as shown in Figure \ref{fig:Ag_recon}. This scalar quantity is the main quantity of interest for reconstructive surgery as it determines the area of the reconstructive skin flap. We define an acceptable error tolerance in the predicted area gain $A^g_{f,pred}$ as a function of $A^g_{f,true}$:

\begin{equation}
T(A^g_{f,pred}) = \max\left(A^g_{\min}, \sqrt{A^g_{f,true}}\right)
\label{eq:rollout_loss}
\end{equation} 

and then take $A^g_{\min}=5cm^2$ as an acceptable minimum error threshold. Our intuition here is that for skin flap reconstruction, the acceptable error depends on the characteristic length scale of the flap. In other words, the length scale that describes the reachability of the flap for skin reconstruction is the important clinical threshold. 
See the Supplementary Information file for more details on how we determined the envelope shown in Figure \ref{fig:Ag_recon}.

The results from Figure \ref{fig:Ag_recon} show that despite low long-horizon losses, Model A systemically under-predicts the deformation, leading to an under-prediction in the final scalar area gain, capturing only 43.7\% of the final points within the acceptable clinical tolerance (note that many validation cases cluster towards $A^g_{f,true}\approx0$.). Model B and Model C also perform poorly on the final scalar integrated area gain, capturing 53.5\% and 33.5\% of the validation cases, respectively. In contrast, Model D achieves promising performance with 90.3\% of the validation cases within the acceptable tolerance band.

Lastly, we tested the rollout prediction time of the Model D architecture on the validation set. The mean evaluation runtime on 4 AMD Epyc Milan processor cores was $0.151\pm0.0341$ seconds which denotes the per-simulation inference time, including model/setup overhead and rollout through the simulation horizon, with final predicted deformation and growth fields formed in memory but not written to disk. This is over $20,000 \times$ faster than the corresponding Abaqus simulations used to generate this data on the same hardware.

\section{Discussion}

To our knowledge, the use of state-dependent feature feedback to stabilize long-horizon predictions in Neural ODE ROMs has not been previously explored. Neural ODEs are inherently limited by their reliance on the initial condition, which prevents incorporation of new information during rollout  \cite{kidger2020neuralcde}.
Neural   controlled differential equations address this limitation by making the latent evolution conditioned on an external control path. In contrast, our approach restores approximate Markovian structure using feature feedback derived from the model’s own latent predictions rather than external inputs. A common idea for improving long-term stability in Neural ODE models by increasing the Markovian nature of the learned dynamics is \textit{memory closure} which involves the addition of a memory term to account for the effect of unresolved dynamics on resolved dynamics, formalized by the Mori-Zwanzig formalism \cite{Mori_MZ_formalism, Zwanzig_MZ_formalism}. This memory term can be approximated using data-driven models such as recurrent neural networks \cite{RNN_mem_closuse}. Subsequent work has incorporated RNN-based memory closures directly into projection-based ROMs, using LSTM networks to approximate the memory integral from short histories of the reduced coefficients \cite{Mem_closure_LSTM}.

The stability of Neural ODE models over long horizons has also been investigated using techniques from feedback control systems.
Recent work has incorporated classical observer design principles into Neural ODEs, embedding Luenberger and Kazantzis–Kravaris–Luenberger structures to promote stability and convergence of the learned latent dynamics \cite{NODE_state_observers}. Related ideas can be found in Koopman-based latent dynamical models, where recent work has addressed long-horizon model drift using projection-based re-encoding mechanisms that activate when drift is detected \cite{Koopman_active_reprojection}, which serves as an active correction strategy applied during rollout. 

However, unlike Mori-Zwanzig-inspired memory-closure approaches that explicitly model history via learned representations or feedback approaches that leverage access to partial ground truth measurements, we attempt to increase the Markovian nature of the learned dynamics through a feature feedback strategy that uses decoded-field features built from the model's own latent prediction without knowledge of the ground truth. 
In other words, the learned dynamics are conditioned on quantities that encode useful structure reconstructed from the model's own latent state. Importantly, however, growth is not treated as a static input: it is evolved externally through a physics-based integration step and fed back into the model, effectively augmenting the system state with physically-consistent information at each time step. 

An interesting result of this augmentation is that Model D performs similarly well even in regimes where negligible growth occurs. This suggests that the benefit of the proposed framework is not solely attributable to the magnitude of growth feedback, but rather to the structured coupling between the learned dynamics and the auxiliary growth state, which regularizes the model even when growth accumulation is inactive.

We believe that our feature feedback approach may hint at a useful insight in learned dynamical ROMs, which is that while the latent POD state may be sufficient to accurately reconstruct the solution, it may not be the most efficient representation for data-driven approaches such as Neural ODEs to learn the conditional dynamics. In our work, the linear basis for the displacement captures 99.99\% of the energy in the training snapshots, yet knowledge of these latent coefficients alone (Model A) is insufficient for accurate long-horizon prediction. The model benefits greatly from learned features that re-encode information about the state in a way that allows the model to more easily learn the long-horizon dynamics (Model D). This observation is consistent with a growing body of work demonstrating that, although a state representation may be sufficient for reconstruction, alternative feature representations can significantly improve the learnability of the underlying dynamics \cite{Noelia_Feature_Operators}. 

The success of this approach in the TE setting can be attributed to two key properties of the underlying physics. First, the problem is quasi-static, eliminating inertial effects and reducing temporal dependencies. Second, tissue growth is cumulative and irreversible, meaning that the relevant history of the system is encoded in the current state. Thus, the displacement dynamics appear to be approximately Markovian when conditioned on the current displacement, growth state, and volume controller feedback terms. This suggests that, in this setting, explicit memory modeling may be unnecessary if the state representation captures the relevant accumulated effects. 

Future work is needed to investigate whether similar feature-based approaches can improve data-driven latent dynamical models in more general systems, particularly those with significant unresolved dynamics that typically require memory closure.

Our Neural ODE ROM is also novel in the way that it implicitly handles the contact dynamics. Contact in ROMs is difficult because contact pressure often does not admit a low-rank structure \cite{Contact_pressure_low_rank}. Approaches to account for this difficulty remain an active topic of research, with recent methods focusing on careful handling via mortar tied-contact constraints \cite{Mortar_based_contact} or retaining the full-order representation of the contact interface while reducing only the interior degrees of freedom \cite{Full_order_contact_ROM}. Indeed, current ROMs in biomechanics often simplify contact by using force boundary conditions \cite{Shah2024ROMStent} or by using analytical models such as sphere-to-plane contact \cite{Foot_ROM_contact}. In our case, the expander-skin contact is very nonlinear with a highly evolving contact area as the expander inflates and its shape changes, making it an especially difficult challenge for current ROM approaches. Fortunately, the expansion process is kinematically driven by the expander volume. In simulation, this volume is controlled via a PID feedback controller, meaning the volume dynamics are almost completely Markovian given only the current volume, volume error and the error history. This allows us to exploit the control-driven kinematics to avoid explicit contact modeling entirely since the corresponding displacement dynamics can be accurately estimated from the driving volume dynamics. 

This work contributes to a gap in ROMs of growth and remodeling (G\&R) in biomechanics. Existing reduced-order and surrogate approaches for G\&R  typically approximate growth through direct interpolation of simulation snapshots or regression of input–output data, rather than learning the evolution of the system state itself \cite{Han_TE_MultiGP, AAA_Multi_GP, reiber2025podrbf_rom_hip}. In contrast, we propose a ROM framework based on a continuous-time latent dynamical formulation, in which the evolution of the system state is learned through a Neural ODE, augmented with physics-informed feature feedback to improve closure of the reduced dynamics. To our knowledge, this represents one of the first applications of a learned latent dynamical ROM that can reconstruct full field G\&R response in biological tissue. 

Notably, we do not reduce the growth model itself; instead, tissue growth is evolved explicitly at the high-fidelity level because it is a linear problem in the case of explicit time integration (even if the right hand side of the growth ODE is nonlinear). Furthermore, the  G\&R model is also spatially decoupled (it does not involve gradients of the growth field) and thus it can be solved in parallel, allowing for high computational efficiency. This suggests that selectively retaining inexpensive mechanistic components can improve closure and stability of ROMs.

Lastly, our ROM shows great promise as a potential rapid surrogate for deformation and growth prediction in TE. It achieves highly accurate deformation predictions over a wide range of expander volumes $V_f\in[100,700]~mL$ which contain the vast majority of clinical final expander volumes which are typically only around $400~mL$ \cite{Median_expd_volume}. The learned reduced-order model achieves over four orders of magnitude speedup relative to high-fidelity finite element simulations performed in Abaqus on the same hardware. Additionally, multiple instances of the model can easily be run in parallel on a laptop. Accuracy on the predicted final scalar area gain is also reasonably high, with 90\% of the validation cases within the clinical tolerance band. We are investigating further improvements, including refining the latent basis, as well as incorporating patient-specific geometric variability to advance toward a clinically useful predictive simulator.

\section{Conclusions}\label{sec13}

In this work, we show that flexible CNN-based closed-loop feedback reduces long-horizon errors in NODE ROMs of skin deformation and growth. The CNN-based feedback model shows low errors in displacement reconstruction across a wide range of expander volumes and simulation  parameters spanning material properties, biological parameters, and boundary conditions. Model prediction time is very fast. The full autoregressive prediction rollout takes about $0.2$ seconds, which is over $20,000\times$ faster than the corresponding Abaqus simulations used to generate the data. More broadly, these results suggest that augmenting reduced-order states with structured, physically-informed features can significantly improve the learnability and long-horizon stability of latent dynamical models. 
Such a framework, given additional data and refinement, is promising for real-time predictive simulation of TE. Future work will investigate geometry variation and refinements to achieve higher accuracy in order to further pave the way for clinical use of real-time predictive digital twins.

\backmatter

\bmhead{Supplementary information}

All code is available at \url{https://github.com/joel-laudo/TE-skin-NODE-ROM}

\bmhead{Acknowledgements}

This work was supported by NIAMS, USA award R01AR074525.

\bmhead{Declarations}

The authors have no conflicts of interest to declare.


\end{document}